\documentclass[fleqn,10pt]{wlscirep}
\usepackage[utf8]{inputenc}
\usepackage[T1]{fontenc}
\usepackage{setspace}
\title{Variations in stability revealed by temporal asymmetries in contraction of phase space flow}
\author[1,*]{Zachary C Williams}
\author[2]{Dylan E McNamara}
\affil[1]{Nicholas School of the Environment, Duke University, Durham, NC}
\affil[2]{Department of Physics and Physical Oceanography, University of North Carolina, Wilmington, NC}
\affil[*]{zachary.c.williams@duke.edu}

\begin{abstract} 
Empirical diagnosis of stability has received considerable attention, mostly focused on variance metrics for early warning signals of abrupt system change. Despite this, the theoretical foundation and application has been limited to relatively simple system changes such as bifurcating fixed points where variability is extrinsic to the steady state. There is currently no foundation and associated metric for empirically exploring stability in wide ranging systems that contain variability in both internal steady state dynamics and in response to external perturbations. Utilizing connections between stability, dissipation, and phase space flow, we show that stability correlates with temporal asymmetry in a measure of phase space flow contraction. Our method is general as it reveals stability variation independent of assumptions about the nature of system variability or attractor shape. After showing efficacy in a variety of model systems, we apply our technique for measuring stability to monthly returns of the S\&P 500 index in the time periods surrounding the global stock market crash of October 1987. Market stability is shown to be higher in the several years preceding and subsequent to the 1987 market crash. We anticipate our technique will have wide applicability in climate, ecological, financial, and social systems where stability is a pressing concern.
\end{abstract}
\begin{document}
\onehalfspacing
\flushbottom
\maketitle
\thispagestyle{empty}

\section*{Introduction}
Comparing stability across systems or forecasting a change in stability when underlying dynamical equations are not known is a central challenge throughout science \cite{holling1973resilience} with far reaching societal relevance. Despite this, there is a lack of an agreed upon interpretation of stability and how it is measured empirically. This is likely due to stability being considered across many disciplines and in a broad array of systems, from simple bifurcating population models to climate models with many nonlinear, interacting parts. We present a coherent and unified tool set for gaining insight into stability based on nonlinear dynamical systems theory, which contains the theoretical apparatus to understand stability in a wide range of contexts.

When a system is nonlinear and dissipative, all trajectories in phase space converge toward a subset of the space called an attractor. The basin of attraction is composed of all states which, through the action of dissipation, eventually lead to the attractor subset of the phase space. Stability in this setting is conditioned on two properties of a system’s phase space; the basin of attraction and dissipation \cite{nicolis1995introduction}. Stability decreases when either the attractor basin range is diminished relative to the size of external system perturbations or the amount of dissipation in the dynamics is reduced. 

Most previous work exploring stability use metrics that capture time series variations to provide a warning of looming change in a system's attractor \cite{scheffer2009early,dakos2015resilience,dakos2017elevated}. These methods fall under the umbrella of ``early warning signals'' for so called critical thresholds. The most prominent of these techniques is referred to as critical slowing down (CSD) \cite{dakos2015resilience}. CSD indicators such as increasing autocorrelation and variance in state variables tend to rise for some systems which presages a critical transition (or tipping point) \cite{van2007slow}. A built in assumption for this technique is that the dynamics are dominated by the return of a system after a perturbation, and that internal system variability remains constant as stability changes. This assumption breaks down for even modest increases in system complexity where variance and autocorrelation can be tied to intrinsic system dynamics and many previous studies have detailed examples where the CSD metrics do not provide insight into system stability \cite{wagner2015false,menck2013basin,karnatak2017early,boettiger2012early,ashwin2012tipping}.

A more recent effort to quantify stability uses time series of multiple ecological species interacting in a network \cite{ushio2018fluctuating}. This novel approach uses Convergent Cross Mapping (CCM) to identify coupled species and then builds a linear prediction model from the reconstructed phase space. While this improves upon previous efforts that focus on simple systems, the appeal to linearized stability metrics keeps the focus of the analysis on a system's return from small perturbations in a linear setting. Said another way, the full scope of potential sources of nonlinear variability and their relation to a system's stability is neglected.

A critical component of this study is how dissipation, revealed in a system's phase space behavior, is manifested on and near the attractor and how it can be revealed using only time series data of the state variables. In this way, we are not assuming a priori that stability is only revealed in how a system responds to perturbations nor are we assuming any particular type of attractor change such as a simple fixed point bifurcation. Our approach is much more general. 

Dissipation arises when differences in state variables are diffusively damped, mixed, or reduced in the phase space and as such it is directly related to the global phase space volume contraction rate and is inversely related to time of decay to an attractor \cite{nicolis1995introduction} (hereafter referred to as volume contraction rate). Consequently, systems with more dissipation are more stable; as a system is drawn more rapidly toward the attractor, state trajectories are more likely than not to stay near to the attractor in the future. Measuring the decay time to the attractor is difficult as it is rare that an observer can conclude precisely when a system is outside its attractor. Additionally, dissipation is not trivial to measure when the system is inside the attractor. Consider the simple cases of a fixed point and limit cycle. For these systems, the evolution appears conservative on the attractor because there is zero net convergence and divergence. The dissipation in these cases is removing energy injected from outside the system (in the form of forcing), and so measuring dissipation is difficult because states are no longer converging on the attractor. In the more complicated setting of strange attractors, convergence and divergence occur simultaneously and heterogeneously throughout the attractor \cite{abarbanel1991variation,norwood2013lyapunov}, in contrast to fixed points or limit cycles. The divergence is related to sensitivity to initial conditions, and convergence (dissipation) acts to keep the systems constrained into a fixed attractor volume. Despite these apparent difficulties in measuring dissipation, we put forth an empirical technique and associated metric that provides a direct correlation to the amount of dissipation in a system. Dissipation as referred to throughout this manuscript is synonymous with phase space volume contraction, e.g. higher rate of dissipation in system dynamics implies higher volume contraction rate in the phase space, and hence more stability.

The classic way to measure the tendency to both expand and contract in phase space is by determining the Lyapunov exponents (LE) \cite{brown1991computing,wolf1985determining}. A dissipative nonlinear dynamical system with $k$-degrees of freedom has $k$ global Lyapunov exponents. These global LEs contain at least one positive (nonlinearity) and one negative (dissipation) exponent and their sum is equal to the average flow convergence rate in the phase space \cite{nicolis1995introduction}. If the spectrum of LE are measured locally in phase space along a trajectory and over some finite time horizon $L$, then they are a function of position along the trajectory $\vec{x}(t)$ and $L$, specified as $\Lambda(\vec{x}(t),L)$. Consider two system states, $\vec{x_i(t)}$ and $\vec{x_j(t)}$, that are initially near neighbors on the attractor at $t$; that is $\left\lVert \vec{x}_i(t)-\vec{x}_j(t)\right\rVert=d_{i,j}(t)\ll1$. After $L$ time has passed, the distance between these points will have grown (or decayed) exponentially:
\begin{align}
    \begin{split}
  d_{i,j}(t+L) &=\left\lVert \vec{x}_i(t+L)-\vec{x}_j(t+L)\right\rVert\approx d_{i,j}(t) e^{L\Lambda(\vec{x}_i(t),L)}\\
      \label{Eq:1}
    \end{split}
\end{align}
where $\Lambda$ is commonly referred to as either the first (i.e. largest, maximal) local LE or the error growth rate, depending upon the direction of initial phase space separation. The phase space average (over all $t$) of the local LEs converges to a single invariant set called the global Lyapunov spectrum. In theory, if the spectrum of global LEs can be determined, then the volume contraction rate, and therefore dissipation rate and stability, can be directly inferred.

In time series applications where data is usually available for only one of many degrees of freedom, the global LEs can be obtained by first invoking Taken’s embedding theorem to reconstruct the attractor \cite{takens1981detecting,abarbanel1993analysis,wolf1985determining}. With the attractor reconstructed, estimates of local LEs may be made. The global LE are then obtained by averaging the local LEs over the entire phase space. However, only the largest (positive) global LE is considered a reliable estimate \cite{kantz2004nonlinear,sugihara1990nonlinear}, and so the volume contraction rate is not reliably known. 

\section*{Phase Space Stability Technique}
In empirical settings, one rarely has access to all perturbation directions in phase space. Additionally, the distribution of local LEs around an attractor is typically not Gaussian \cite{abarbanel1993analysis}. It is simply unfeasible in many circumstances to numerically quantify the dissipation rate (volume contraction rate) via the sum of the globally averaged local LEs. Instead we find a measure which correlates with dissipation rate that is based on the growth rate of the globally averaged local separation distances $\langle d_{t+L}/d_t\rangle$: 
\begin{align}
    \begin{split}
    \lambda(L) &=\frac{1}{L}log\langle\frac{d_{i,j}(t+L)}{d_{i,j}(t)}\rangle
    \label{Eq:2}
    \end{split}
\end{align}
We interpret $\lambda$ as the global average trajectory divergence rate given as a function of some finite time horizon $L$. The brackets $\langle\cdot\rangle$ refer to taking a phase space average, that is, evaluating for all points $i$ along some trajectory. As we will show numerically for several model nonlinear dissipative systems, there is a temporal asymmetry in $\lambda$ such that upon evaluation of equation \ref{Eq:2} forward in time ($\lambda^+$) and backward in time ($\lambda^-$) and differenced, has a maximum (symbolized herein $\Delta\lambda$) which correlates with the dissipation rate and hence stability.

To calculate $\Delta\lambda$ numerically, we first partition a time series in to two sets that are presumed to mostly lie on the attractor. One partition becomes a test set, and the other becomes a library set from which near neighbor trajectories are chosen. For each point in the test set, the nearest neighbor is located and a time series of separation distances is catalogued. Each time series is normalized with respect to the initial separation distance, then averaged resulting in the average separation distance as a function of $L$. Finally the logarithm of the average separation distance is divided by the time since the initial separation ($L$). This gives the forward divergence rate $\lambda^+$ as a function of $L$ (Eq. 2). The procedure for calculating the backward divergence rate $\lambda^-$ is the same as the forward, but with time reversed i.e $d_{t-L}=\left\lVert\vec{x}_i(t-L)-\vec{x}_j(t-L)\right\rVert$. Finally, our metric $\Delta\lambda$ is obtained by taking the maximum of the difference between $\lambda^+$ and $\lambda^-$
\begin{align}
    \begin{split}
    \Delta\lambda=max\{\lambda^-(L)-\lambda^+(L)\}.
    \end{split}
    \label{Eq:3}
\end{align}
which we demonstrate is correlated with the system’s dissipation rate yielding a quantitative measure of dissipation and therefore stability.

\section*{Results}
 The efficacy of our phase space stability technique is demonstrated by application to the canonical Lorenz system\cite{lorenz1963deterministic}. Utility is then demonstrated in reconstructed phase space of the Lorenz system, the Lorenz system with multiplicative noise, and the R\"{o}ssler system with multiplicative noise. Finally we present an application to the financial market crash of October 1987, an event that is widely reported as having resulted from a loss of internal system stability.

\subsection*{Application to the Lorenz system}
\begin{figure}[!ht]
    \centering
    \includegraphics[width=2 in]{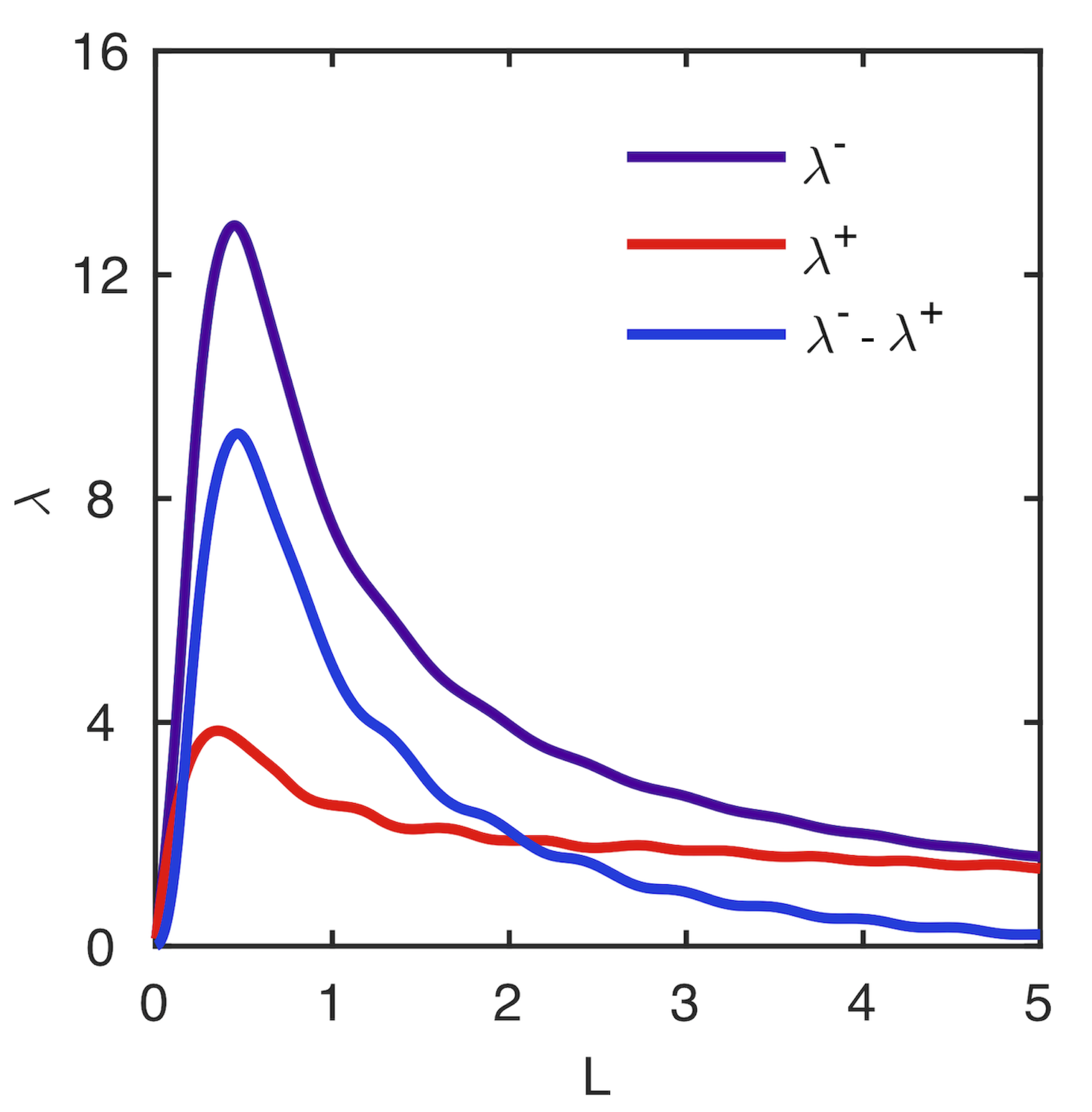}
    \caption{The average trajectory divergence rate as a function of time horizon ($\lambda(L)$) for Lorenz system with parameters $r=45$, $b=8/3$, and $s=20$. The forward time divergence rate ($\lambda^+$) is red and the backward time divergence rate ($\lambda^-$) is indigo. The difference between backward and forward divergence rates is blue. Here the phase space volume contraction rate is approximately -23.6.}
    \label{fig:1}
\end{figure}

The phase space stability technique is demonstrated examined first with the Lorenz system. Details of numerical method are found in the Methods section. Library and test sets are constructed from the three system variables $x,y$ and $z$, and the parameter values used are $r=45, b=8/3$ and $s=20$. The forward divergence rate $\lambda^+$ as a function of $L$ is the red curve in figure \ref{fig:1}. When time is reversed, the divergence rate ($\lambda^-$) results in the indigo curve in figure \ref{fig:1}. The difference between backward and forward divergence rates is the blue curve in figure \ref{fig:1}, which typically peaks at intermediate values of $L$ before vanishing as $L$ grows large. We note that in some cases there is no observable peak in $\lambda^+$ or $\lambda^-$, such that divergence rates are large for small $L$ and decreasing with increasing $L$. In these cases, there is still a clear peak in their difference. 

\begin{figure}[!ht]
    \centering
    \includegraphics[width = 4.5in]{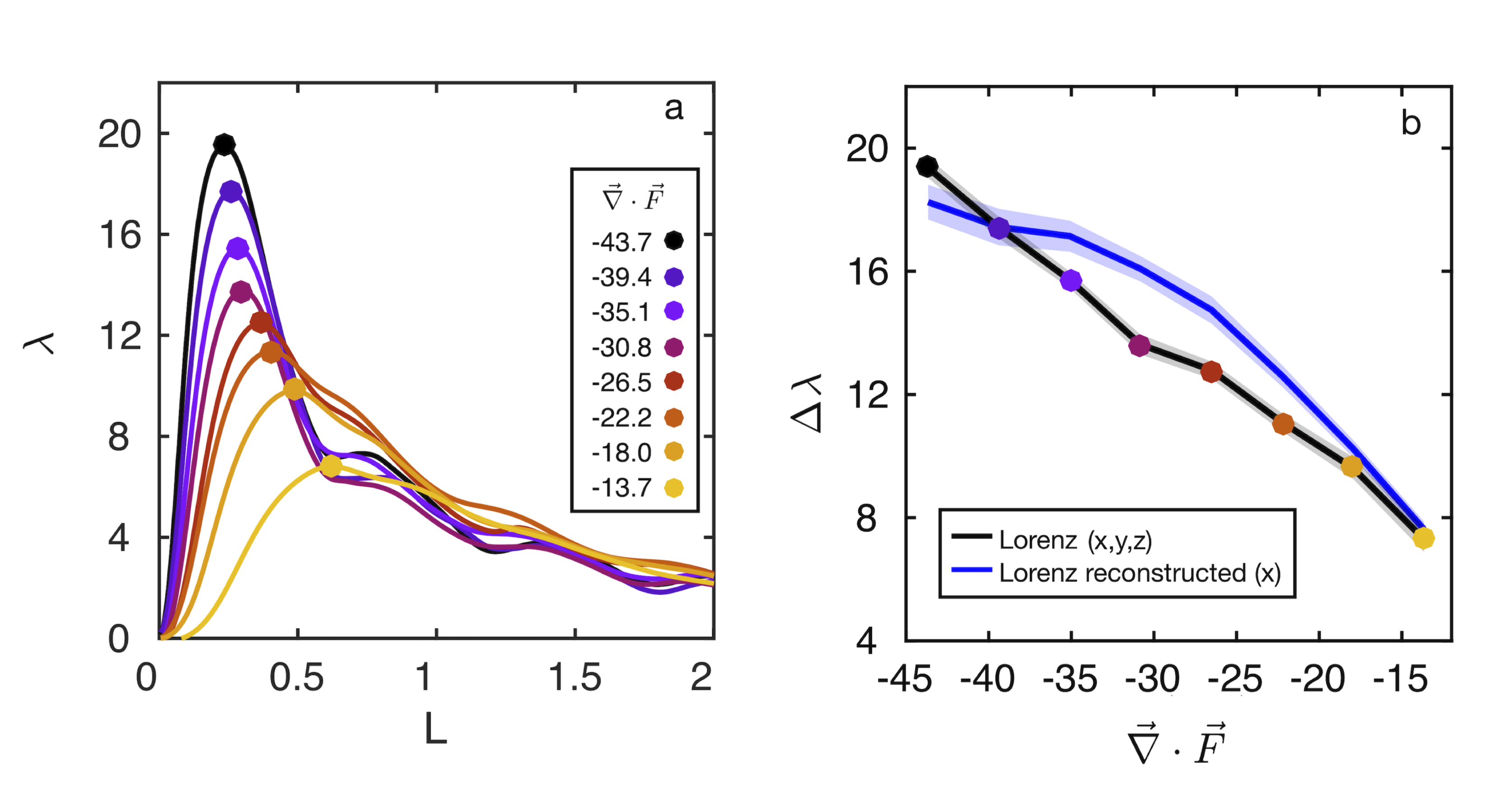}
    \caption{Variation of the stability metric with volume contraction for the Lorenz system. Panel (a) plots the difference between backward and forward divergence rates as a function of $L$ based on the full Lorenz attractor where line color corresponds to the value of phase space volume contraction rate $\vec{\nabla}\cdot\vec{F}$ which is varied through the dissipative control parameter $s$. Panel (b) shows $\Delta\lambda$ as a function of volume contraction rate for the full Lorenz attractor (black), and the reconstructed attractor (blue). The standard deviation of $\Delta\lambda$ is indicated by the lighter shade.}
    \label{fig:2}
\end{figure}

Next we show how $\Delta\lambda$ varies with the phase space volume contraction rate in the Lorenz system, where the contraction rate is controlled via the parameter $s$, that is $\vec{\nabla}\cdot\vec{F}(s)$. The parameter $s$ is varied between between 10 and 40, while $r=45$ and $b=8/3$ are fixed. Figure \ref{fig:2}a plots $\lambda^-(L)-\lambda^+(L)$ for 8 values of the contraction rate. A key element of figure \ref{fig:2}a is demonstrated more clearly by the black curve in figure \ref{fig:2}b where the peaks ($\Delta\lambda$) identified in figure \ref{fig:2}a are plotted against volume contraction rate. Each point along the black curve in figure \ref{fig:2}b corresponds to the ensemble mean of $\Delta\lambda$ pertaining to 100 repeated solutions of the Lorenz system with random initial conditions, and the gray shaded bounds correspond to the ensemble standard deviation. A statistically monotonic relationship between $\Delta\lambda$ and contraction rate is observed such that as the volume contraction rate increases (or stability increases), the magnitude of the asymmetry between forward and backward divergence rates is larger. An analytical representation of the non-isotropic flow divergence on the attractor and its variation with control parameters (here $s$) related to volume contraction escapes us and as far as we can tell, has escaped the community. Thus, as a first step, we are only revealing the correlation between our measure of the asymmetry $\Delta\lambda$ and the volume contraction rate. Additionally, if contraction rate were instead varied as a function of $b$, similar results are obtained, albeit there is an altogether smaller range of bounded or numerically stable solutions (see Supplementary Figure S1 in the accompanying Supplementary Information file).

To test the efficacy of the $\Delta\lambda$ metric in correlating with volume contraction rate for a reconstructed attractor, we used reconstructed the phase space of the Lorenz system based on time series of the $x$-variable over the same range of control parameters used to generate the black curve in figure \ref{fig:2}b. The blue curve in figure \ref{fig:2}b displays $\Delta\lambda$ as a function of volume contraction rate for 100 simulations per choice of $s$. Even in the reconstructed phase space, a statistically monotonic relationship between $\Delta\lambda$ and volume contraction rate is observed. We also observed this relationship holds in the $2^{nd}$ moment and higher order distributions of the largest local LE and error growth rates (as defined in \cite{abarbanel1991variation,trevisan1995transient} respectively). 

\subsection*{Application to Stochastic Nonlinear Dissipative Systems}
While the Lorenz system is useful for illustrative purposes, it is rare that an empirical investigation will find such a smooth, low dimensional dynamical system. More commonly, irregular system behavior is driven by both low dimensional nonlinearity and the influence of noise, which can be interpreted as a connection to a large reservoir of unmeasured degrees of freedom. To explore the efficacy of our metric in measuring stability when noise is dynamically embedded into the low dimensional nonlinear dynamics, we investigate the Lorenz system and R\"{o}ssler system with multiplicative (state-dependent) Guassian noise.

\begin{figure}[!ht]
    \centering
    \includegraphics[width=5 in]{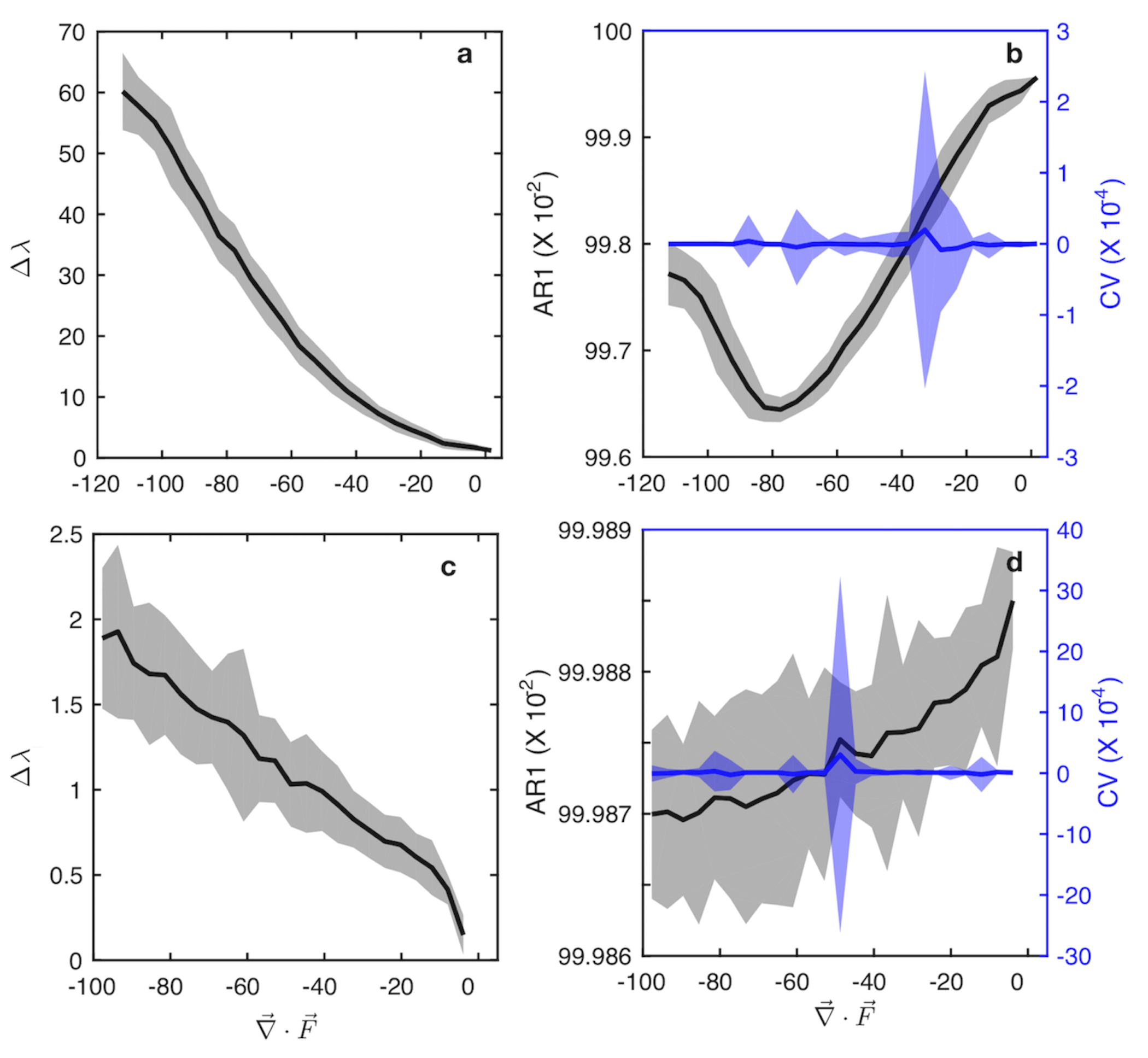}
    \caption{$\Delta\lambda$ as a function of phase volume contraction rate for the reconstructed stochastic Lorenz attractor with linear multiplicative noise (a) and for the reconstructed stochastic R\"{o}ssler attractor with multiplicative noise (c). Panels (b) and (d) compare the two most common critical slowing indicators: the Autoregressive lag-1 (AR1) parameter and coefficient of variation (CV) corresponding the Lorenz (b) and R\"{o}ssler (d) systems with multiplicative noise respectively, given as a function of phase volume contraction rate.}
    \label{fig:3} 
\end{figure}

The variation of $\Delta\lambda$ as a function of contraction rate for the reconstructed phase spaces of the stochastic Lorenz and R\"{o}ssler systems is shown in figures \ref{fig:3}a and \ref{fig:3}c respectively. The shaded grey regions convey the ensemble mean and standard deviation from 100 repeated solutions of both systems with random initial conditions. In both systems, $\Delta\lambda$ increases as the volume contraction rate increases.

For systems that approach a simple bifurcation when varying a control parameter, the coefficient of variation ($CV$) and auto-regressive lag-1 coefficient ($AR(1)$) have been demonstrated to increase, providing a kind of early warning signal of the impending bifurcation. These measures are commonly referred to as critical slowing down indicators. In figures \ref{fig:3}b and 3d, we evaluate $CV$ and $AR(1)$ for the stochastic Lorenz and R\"{o}ssler systems over the same time series ensembles used to generate figures \ref{fig:3}a and c. For the stochastic Lorenz system, the $AR(1)$ coefficient (black line, figure \ref{fig:3}b) decreases over a range of increasing volume contraction rates, but then changes direction, indicating it is not reflecting phase space stability as explored here. It is also interesting to note that changes in $AR(1)$ are very small. The coefficient of variation $CV$ is the blue line in figures \ref{fig:3}b and does not bear any relation to the volume contraction rate. For the stochastic R\"{o}ssler system, the $AR(1)$ coefficient (black line, figure \ref{fig:3}d) tracks with volume contraction rate, however changes again occur at the third decimal and the uncertainty bounds are so large it renders any distinction moot. The coefficient of variation is the blue line in figures \ref{fig:3}d, which also does not bear any relation to the volume contraction rate. Figure \ref{fig:3}b and d demonstrate that the critical slowing down indicators considered here do not relate to stability when the system is more complicated than classic fixed point bifurcations. 

\subsection*{Application to Stock Market Stability}
Empirical analysis tools related to dissipative nonlinear dynamical systems \cite{brock1991nonlinear,zanin2018assessing}, and in some cases specifically attractor reconstruction \cite{berg2015economic,hsieh1991chaos}, have been used in a wide range of economic settings. In this spirit, here we apply our metric to evaluate financial market stability during the 1980s, with specific focus on the October 1987 financial market crash referred to as Black Monday. Financial markets can be considered as complex adaptive systems composed of many heterogeneous interacting agents who process information to form expectations based on exogenous (e.g. news) and endogenous sources (e.g. other agent opinions) with the goal of maximizing stock market investments \cite{arthur1999complexity,hommes2001financial}. Both empirical and theoretical studies show strong support for this dynamical systems conceptualization of economies and markets, for example see \cite{shiller1990speculative,sornette1997large,brock1991nonlinear}. A market crash can result from both exogenous shocks to the economy (e.g. a pandemic) or endogenous dynamics (e.g. speculative bubbles), or some combination therein \cite{johansen2010shocks}. Black Monday was the single largest drop in the history of the S\&P500 and is considered to be entirely the result of internal dynamics, specifically positive feedbacks between speculative and fundamentalist stock traders \cite{sornette1997large,shiller1990speculative}. 

\begin{figure}[!ht]
     \centering
     \includegraphics[width=2.5 in]{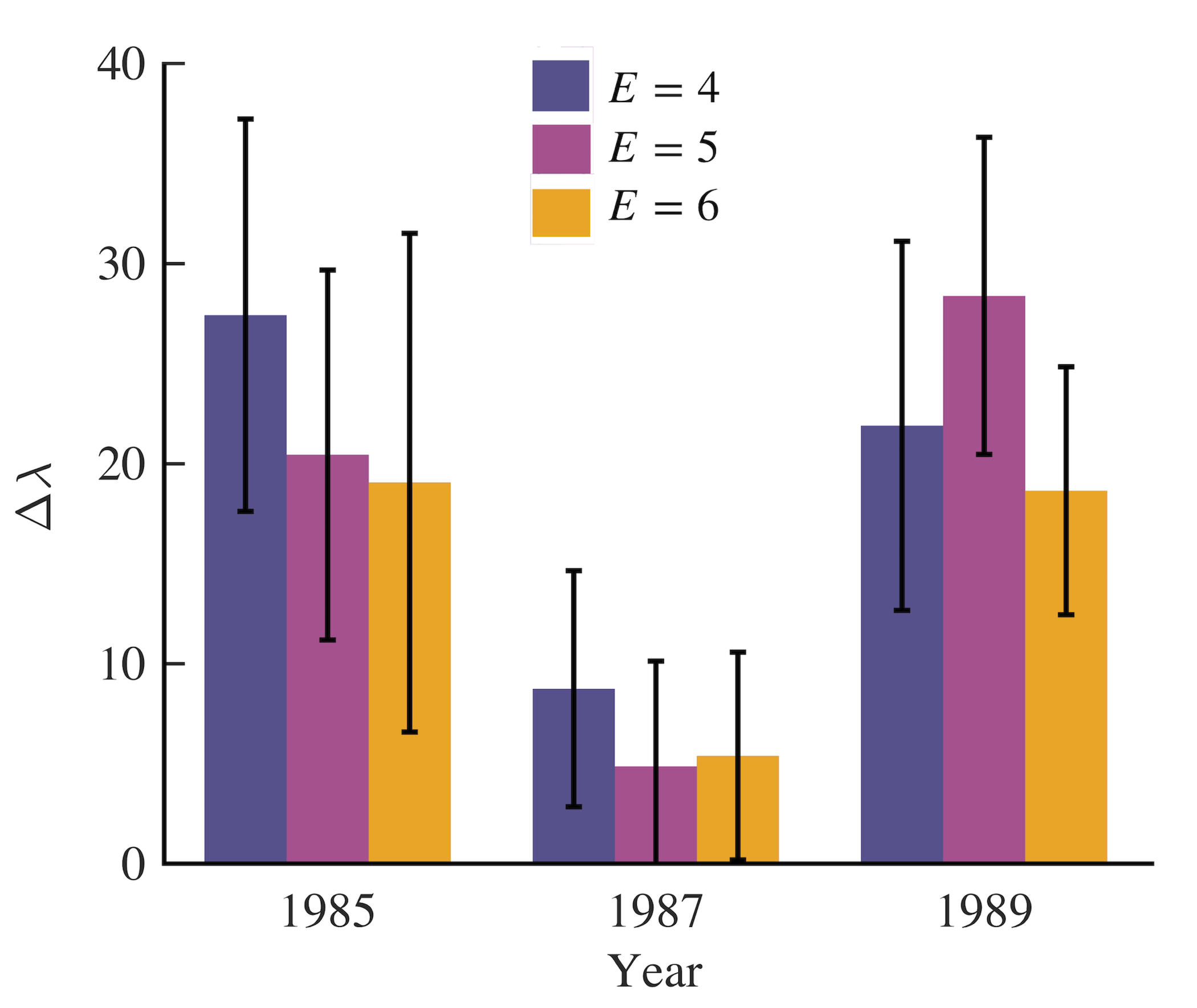}
     \caption{Stability of the S\&P 500 index during three time periods in the mid to late 1980s. $\Delta\lambda$ is shown as a function of time (year) and for three values of phase space embedding dimensions (indicated by color). Error bars correspond to one standard deviation of the distribution of $\Delta\lambda$ calculated over each corresponding time period. These results demonstrate that market stability was significantly lower in the time around the October 1987 global stock market crash, when compared to the years before and after the crash.}
      \label{fig:4}
 \end{figure}

We apply the phase space stability technique to the price return time series of the S\&P500. Figure \ref{fig:4} demonstrates that system stability $\Delta\lambda$ was higher in the years preceding and proceeding the 1987 crash, and was nearly absent in vicinity of the crash. While previous work suggests a suitable embedding dimension ($D_E$)of 5 for the S\&P 500 returns time series, robustness of the result is partially validated by testing $D_E$ of 4,5, and 6. For all three choices of $D_E$, $\Delta\lambda$ is lower in the time period around 1987. 

\section*{Discussion}
We have presented a technique to quantify the phase space stability of nonlinear dissipative systems based on time series observations. The technique is applied in the context of canonical nonlinear and stochastic dynamical systems and we provide application to the S\&P 500 time series that contains a well understood and verifiable financial market instability which occurred globally in 1987. Stability in the context of the present study refers to the internal system stability reflecting interplay of the underlying nonlinear and dissipative dynamic processes. Previous efforts to quantify stability from time series effectively assume stability is constant throughout the attractor and that a change in stability is immediately detectable. For example, Critical Slowing Down indicators would interpret a change in the amplitude or variability of an external stochastic forcing as a change in stability, even while nothing has changed in the internal dynamics. Our technique assesses the context of a perturbation (i.e. the attractor), and therefore would correctly predict no change in stability. This is because our technique quantifies the rate of dissipation across the whole attractor, as opposed to extrapolating local stability properties of the attractor which are well-known to vary heterogeneously throughout the attractor.

Why should we expect an observable temporal asymmetry in the phase space of dissipative nonlinear dynamical systems to exist and why would this be connected to stability? Temporal irreversibility in nonlinear time series analysis has been discussed in literature, typically focusing on the presence of time series irreversibility as an indication of nonlinear determinism \cite{stone1996detecting} and as a reflection of the entropy production by the underlying dynamical processes \cite{zanin2018assessing} \cite{roldan2010estimating}. Relatively recent advances in the theory of nonequilibrium statistical mechanics explicitly connect entropy-production and phase space volume contraction for nonequilibrium systems \cite{daems1999entropy}. Additionally, the Crooks Fluctuation Theorem \cite{crooks1999entropy} \cite{jarzynski1997nonequilibrium} connects the forward and reverse time phase space trajectories to dissipation. Our work builds on these insights by noting that since temporal irreversibility is due to entropy production (dissipation), and entropy production rate is equivalent to the phase space volume contraction rate, then a measurement of time asymmetry should also reflect stability because the volume contraction rate is the sum of the Lyapunov exponents governing the stability of trajectories in phase space.

We hypothesize that backward time divergence is larger than forward time divergence owing to variations in the strength of converging and diverging regions on an attractor \cite{sterk2012predictability,trevisan1995transient} and the action of dissipation in reducing differences in system state. While volume contraction rate in the full space for the Lorenz attractor is constant, the rate of separation between trajectories varies around the attractor as the direction to near neighbors varies throughout the phase space. When choosing points to test for distance spreading, we use nearest neighbors which ensures that we have chosen from regions of strong dissipation and hence relatively strong flow convergence (dissipation reduces state differences). Conversely, when one marches backwards in time from these close neighbor points on the attractor, the flow tends toward divergence. To be clear, this is not true for every point used in the analysis but when averaged around the attractor, the choosing of very near neighbors has provided enough preference to areas of dissipation to reveal a strong time asymmetry. In fact, if one uses neighbors that are far apart to calculate $\Delta\lambda$, the asymmetry vanishes (not shown).

In the context of attractor reconstruction, the technique presented here is subject to the same limitations that have been exhaustively discussed elsewhere, e.g. data length and stationarity \cite{kantz2004nonlinear}. Another limitation is that a particular value of our dissipation metric, $\Delta\lambda$, does not have any obvious connection to the analytically calculated rate of phase space convergence. It is only in comparing $\Delta\lambda$ for similar systems or evaluating $\Delta\lambda$ through time that one gains insight into relative stability. 

The range of potential applications for our dissipation metric is as wide as the range of utility for attractor reconstruction. One realm of application is in model testing. A given numerical model will have measurable and controllable amounts of dissipation. By comparing two simulations with varying amounts of dissipation to a time series from a natural system, one should be able to test a model's ability to simulate the relative stability of the system in question by measuring our metric for the model and natural system. Beyond model testing, particularly provocative opportunities for using our metric include gaining insight into the amount of dissipation and stability and how that has changed over time in increasingly stressed climate, ecological, financial, or social systems. 

\section*{Methods}
\subsection*{Lorenz System}
The Lorenz system is a set of coupled nonlinear ordinary partial differential equations. There are three degrees of freedom, $x,y,$ and $z$, and three constants $s$, $r$, and $b$. Numerical solutions are obtained using a fourth order Runge-Kutta method. The phase volume contraction rate is determined by taking the divergence of the Lorenz system of equations ($\vec{F}$):
\begin{align}
    \begin{split}
    \vec{\nabla}\cdotp\vec{F} &=-s-1-b
    \end{split} 
    \label{Eq:5}
\end{align}
The volume contraction rate is a function of two parameters $s$ and $b$. For simplicity, we take $s$ as the parameter controlling the divergence rate for all results pertaining to the Lorenz system. However we find similar results when taking $b$ as the controlling parameter. 
and the model time step is $\Delta t=5x10^{-3}$

\subsection*{Stochastic Lorenz System}
The Lorenz system with multiplicative noise is a set of Ito stochastic differential equations \cite{dijkstra2013nonlinear}:
\begin{align}
    \begin{split}
    {dx} &=s(y-x)dt+\sigma x dW_t\\
    {dy} &=(rx-y-xz)dt+\sigma y dW_t\\
    {dz} &=(xy-bz)dt+\sigma z dW_t
    \end{split}
    \label{Eq:6}
\end{align}
The term $dW_t$ is the increment of a Wiener process, and $\sigma^2$ is variance. Parameter values for $r$ and $b$ are the same as those from Figure \ref{fig:2}. The volume contraction rate is similar to that in the deterministic Lorenz but with an additional term reflecting the contribution of the multiplicative noise term \cite{geurts2019lyapunov}:
\begin{align}
    \begin{split}
    \vec{\nabla}\cdotp\vec{F} &=-s-1-b+3\sigma\lim_{t\to\infty}\frac{W_t}{t}
    \end{split} 
    \label{Eq:7}
\end{align}
The range of contraction rates in figure \ref{fig:3}a is obtained by varying $s$ between 2 and 100. The noise standard deviation ($\sigma$) is set to 0.2. Numerical solution to equation \ref{Eq:6} is obtained using the Euler-Maruyama method with a time step of $\Delta t=5\times10^{-3}$. The stochastic Lorenz attractor is reconstructed from the $x$ variable using an embedding dimension $D_E=3$ and embedding time lag $\tau=20$. The multiplicative noise term contributes a small random component to the contraction rate (eq. \ref{Eq:6}) and so ensemble results are displayed as binned averages.

\subsection*{Stochastic R\"{o}ssler System}
The stochastic R\"{o}ssler system \cite{rossler1976equation} with multiplicative noise, similar to the stochastic Lorenz system (eq. \ref{Eq:6}), is a set of Ito stochastic differential equations:
\begin{align}
    \begin{split}
    {dx} &=s(y-z)dt+\sigma x dW_t\\
    {dy} &=(x+ay)dt+\sigma y dW_t\\
    {dz} &=(b+xz-cz)dt+\sigma z dW_t
    \end{split}
    \label{Eq:8}
\end{align}
where $a$,$b$, and $c$ are parameters. The average volume contraction rate is: 
\begin{align}
    \begin{split}
    \vec{\nabla}\cdotp\vec{F}=a-c+\bar{x}+3\sigma\lim_{t\to\infty}\frac{W_t}{t}
    \end{split}
    \label{Eq:9}
\end{align}
Results presented in figure \ref{fig:3}c are obtained by varying the dissipative control parameter $c$  between 2 and 100, the noise standard deviation is $\sigma=0.2$, and the fixed parameter are $a=0.1$ and $b=0.3$. Numerical solution to equation \ref{Eq:7} is obtained using the Euler-Maruyama method with a time step of $\Delta t=5\times10^{-3}$. The stochastic R\"{o}ssler attractor is reconstructed based on the $x$ variable using an embedding dimension $D_E=3$ and embedding time lag $\tau=40$. 

\subsection*{Attractor Reconstruction}
The delay-embedding theorem offers a way to recover the complete phase space behavior of a dynamical system from a time series of just one of the system variables. To reconstruct the attractor, a time series, $x(n)$, is embedded into a $d$-dimensional space to form a trajectory composed of vectors ($\vec{y}_n$) whose components are lagged sequences of the original time series:
\begin{align}
    \begin{split}
     \vec{y}_n=[x(n),x(n-\tau),...,x(n-(d-1)\tau)]
    \end{split}
    \label{Eq:10}
\end{align}
where the constants $d$ and $\tau$ are referred to as the embedding dimension and time delay respectively. Critically, the reconstructed attractor $\vec{y}_n$ is identical to the unknown attractor up to a smooth local change of coordinates, and contains all the topological properties of the unknown attractor i.e. system invariants. 

There is an extensive literature on how to appropriately choose the values of $\tau$ and $d$. We take the first minimum of the mutual information \cite{abarbanel1991variation} to determine $\tau$ and the number of degrees of freedom from the originating system as the embedding dimension. When comparing stability between similar systems (as in figures \ref{fig:2},\ref{fig:3}, and \ref{fig:4}, the choice of $tau$ is kept fixed. This is because an optimized prediction horizon is not the objective here. The objective is to detect relative changes in the flow contraction which could be obfuscated by embedding similar systems with widely varying embedding time lags $\tau$. 

\subsection*{S\&P 500 Index Returns Time Series}
Time series for S\&P500 \cite{SP500} returns are based on the adjusted closing prices $P$. The price return at time $t$ over some interval $T$ is:
\begin{align} 
    \begin{split}
    r^T_t=\frac{P_t-P_{t-T}}{P_{t-T}}
    \end{split}
    \label{Eq:11}
\end{align}
For example when $T=1$ then the returns are daily. Since daily returns are very noisy, we analyze monthly returns ($T=20$) for the analysis presented herein. After obtaining monthly returns, the returns time series is divided into 3 groups spanning the years 1984-1986, 1986-1988, and 1988-1990. This way we test the stability preceding, during, and after the Black Monday crash of 1987. Within each group, we calculate $\Delta\lambda$ based on a sliding sliding window library and test set that are each 252 points (1 year). For example, in the grouping spanning 1984 through 1986, the first library set spans 01/03/1984 to 12/28/1984 and the test set spans 01/03/1985 through 12/28/1985 and $\Delta\lambda$ is estimated. This procedure is repeated by advancing to the start and end dates of both the library and test sets by one day, until the end date of the test set reaches 12/28/1986. The same procedure is applied to each group resulting in approximately 252 estimates of $\Delta\lambda$. The mean and standard deviation are presented in Figure \ref{fig:4}. The embedding time lag is obtained from the first minimum in the average mutual information from the returns time series spanning 1980-1990 and is found to be $\tau=10$. Previous studies have suggested an embedding dimension of around 5 for the S\&P500 \cite{peters1991chaotic}, therefore we present results corresponding to embedding dimensions of 4,5, and 6. 
 
\section*{Acknowledgements}
We are grateful to B.T. Werner and S. Singh for insightful conversations and feedback. Support for this project was provided by the National Science Foundation (EAR 1715638).
\bibliography{stability_bib}
\end{document}